# Bayesian DEJD model and detection of asymmetric jumps[*]

Maciej Kostrzewski[†]

July 31, 2018

**Abstract**

News might trigger jump arrivals in financial time series. The "bad" and "good" news seems to have distinct impact. In the research, a double exponential jump distribution is applied to model downward and upward jumps. Bayesian double exponential jump-diffusion model is proposed. Theorems stated in the paper enable estimation of the model's parameters, detection of jumps and analysis of jump frequency. The methodology, founded upon the idea of latent variables, is illustrated with two empirical studies, employing both simulated and real-world data (the KGHM index).

**Keywords:** double exponential jump diffusion model, Kou model, Bernoulli jump-diffusion model, MCMC methods, latent variables

## 1 Introduction

News concerning the companies, macroeconomic releases, cataclysms or wars has a huge impact on prices of shares, derivative securities, yields, commodities etc. ([1]). Markets often react in a spontaneous way on flowing news. The reactions manifest themselves as jumps in time series.

There are models where jumps and small changes of values are considered simultaneously. Examples of such specifications include the jump-diffusion models and their discretizations (e.g. [2], [3], [4], [5], [6], [7], [8], [9], [10], [11], [12], [13], [14], [15], [16], [17], [18], [19], [20], [21], [22], [23]). One of the best known jump-diffusion model is the Merton model ([24]). In the Merton model, the jumps appear at random moments of time governed by the exponential distribution, whereas the number of jumps and their magnitudes are driven by the Poisson process and the normal distribution, respectively. The process of prices is continuous between jumps — just as in the Black-Scholes model ([25]).

It is generally known that the investors' reaction on "bad" and "good" news is different (crashophobia ([26])). In modelling time series it is a common way to account for this by employing distinct distributions for the negative and the positive jumps. An example of such an approach is to apply a double exponential distribution. In this case the negative and the positive jump distributions are exponential with some (distinct) parameters. In the Merton model, jump values are modeled via a normal distribution. However, if we replace the normal distribution with the double exponential one, we get a specification in which the negative and the positive jumps are handled separately. In this paper, I concentrate on discrete version of such constructions.

[*]The research was partially supported by the Polish Ministry of Science and Higher Education. Research Project realized over 2010-2012; Nr N N111 429139.

[†]Cracow University of Economics, Poland; e-mail: kostrzem@uek.krakow.pl

In the jump-diffusion framework, the distribution of logarithmic returns is given by an infinite mixture of normal distributions. In practice, estimation of this model's parameters is conducted for some model approximation given by finite mixtures. The most famous approximation of the Merton model is the Bernoulli jump-diffusion model ([2]), which allows for at most a single jump per a unit of time (e.g. a day). The same idea is applied to the jump-diffusion model with the double exponential jump distribution. Such specification was considered by Kou ([5]) in the context of pricing derivative securities and it is known as the Kou model. Moreover, it was analysed by Ramezani and Zeng ([14]). This model is a special case of the Pareto-Beta jump-diffusion specification proposed by Ramezani and Zeng ([4]), where two Poisson processes govern the arrival rate of "bad" and "good" information.

In this paper, I consider discretization of the double exponential jump-diffusion model, called the DEJD model. It is equivalent (under an appropriate parametrizations) to the model considered by [5], [14] and [21]. Under the DEJD specification, a single Bernoulli process controls jumps arrivals in returns, wheras the magnitudes of the upward and the downward jumps are generated by the double exponential distribution. The aim of the paper is to develop a Bayesian framework for the DEJD model under (some) proper priors. The idea underlying the statistical model is based on introducing latent variables. Moreover, I give a recipe how to conduct the Bayesian inference in practice, providing schemes of relevant numerical algorithms. Frame and Ramezani ([21]) proposed the Bayesian specification for the equivalent mathematical model. They considered non-informative prior specifications with an exception of the jump intensity parameter. Bayesian framework for models with normal jump values is considered by Rifo and Torres ([17]), Lin and Huang ([6]) and Kostrzewski ([22], [23]). The Merton model, Kou model and DEJD model are used in portfolio choice, pricing derivative securities and risk analysis. From a practical point of view, a reliable method for estimation of this model is of utmost importance. Finally, let me clarify that I am preoccupied with detecting jumps rather than relating them with, e.g., macroeconomic releases. The latter has been attempted by, e.g., [10] and [20].

The remainder of the paper is organized as follows. In Section 1, the theoretical details of the DEJD model are presented. The Bayesian DEJD model is defined in Section 2. Moreover, I propose numerical algorithms based on MCMC methods, which make Bayesian inference possible to apply. In Section 3, empirical results are reported. First, simulated data and then real-world data are considered. The paper ends with some brief conclusions. The proofs of the proposed theorems are provided in the Appendix.

## 2 The DEJD model

Consider a standard Wiener process $W = (W_t)_{t\geq 0}$, a Poisson process $N = (N_t)_{t\geq 0}$ with the intensity $\lambda > 0$, and independent random variables $Q = $

$(Q_j)_{j\geq 1}$ such that $Q_j$ has a double exponential distribution with density

$$f_{Q_j}(x) = p_D \eta_D \exp(\eta_D x) \mathbb{I}_{(-\infty,0)}(x) + p_U \eta_U \exp(-\eta_U x) \mathbb{I}_{[0,\infty)}(x), \quad (1)$$

where $\eta_U > 0, \eta_D > 0$. Let us assume that $W$, $N$ and $Q$ are independent. Finally, $S = (S_t)_{t\geq 0}$ denotes the price process of some risky asset.

The logarithm of $S$ is governed by a jump-diffusion process that constitutes the solution of the equation:

$$d(\ln S_t) = \left(\mu - \frac{1}{2}\sigma^2\right) dt + \sigma dW_t + Q dN_t.$$

It might be shown that

$$S_t = S_0 \exp\left(\left(\mu - \frac{1}{2}\sigma^2\right) t + \sigma W_t + \sum_{i=1}^{N_t} Q_i\right),$$

$$\ln\left(\frac{S_{t+\Delta}}{S_t}\right) = \left(\mu - \frac{1}{2}\sigma^2\right) \Delta + \sigma(W_{t+\Delta} - W_t) + \sum_{i=N_t+1}^{N_{t+\Delta}} Q_i, \Delta > 0.$$

The process is built of two components: the (pure) diffusion part,

$$\left(\mu - \frac{1}{2}\sigma^2\right) \Delta + \sigma(W_{t+\Delta} - W_t),$$

represent continuous variations, wheras the (pure) jump component,

$$\sum_{i=N_t+1}^{N_{t+\Delta}} Q_i,$$

reflects abnormal (extreme) movements in returns. There are three sources of randomness: $W$, $N$ and $Q$, affecting $S$. The (continuous) price behaviour between jumps is described by the geometric Brownian motion, $W$. The arrival rate of jumps is described by the Poisson process, $N$, and the jump magnitudes - by $Q$. The process $S$ depends on six unknown parameters: $\mu, \sigma, \lambda, p_U, \eta_U$ and $\eta_D$.

Before the Bayesian framework for the DEJD model is discussed (see Section 3), we provide some basics underlying the very specification of the model in question. The density of logarithmic rates of return, $\ln\left(\frac{S_{t+\Delta}}{S_t}\right)$, is an infinite mixture:

$$\sum_{k=0}^{\infty} \exp(-\lambda \Delta) \frac{(\lambda \Delta)^k}{k!} f_k(x), \quad (2)$$

where $f_k$ are some densities. Because the series given by (2) is infinite, the density is intractable. Consider an approximation

$$\sum_{k=0}^{\infty} \exp(-\lambda \Delta) \frac{(\lambda \Delta)^k}{k!} f_k \approx \sum_{k=0}^{M} \exp(-\lambda \Delta) \frac{(\lambda \Delta)^k}{k!} f_k \quad (3)$$

for some $M > 0$. The approximation restricts the number of jumps over any time interval $\Delta$ to $M$. The case of $M = 0$ indicates no jumps over interval $\Delta$.

Let us restrict further considerations to the discrete time framework. Time series $(x_1, x_2, ...)$ for $x_i = \ln\left(\frac{S_{t_{i+1}}}{S_{t_i}}\right)$ is observed at $(t_1, t_2, ...)$. Moreover, $\Delta \equiv t_{i+1} - t_i > 0$ is a fixed time interval between following observations. Denote the vector of parameters as $\theta = (\mu, \sigma, \lambda, p_U, \eta_U, \eta_D)$, where $\theta \in \mathbb{R} \times (0, \infty) \times (0, \infty) \times (0, 1) \times (0, \infty) \times (0, \infty)$. If we normalize the approximation given by (3), we obtain the conditional data density (given the parameters, $\theta$):

$$p(x|\theta; M) = \sum_{k=0}^{M} w_k f_k(x), \qquad (4)$$

where $w_k = \frac{(\lambda\Delta)^k}{k!} \left[\sum_{j=0}^{M} \frac{(\lambda\Delta)^j}{j!}\right]^{-1}$ and $f_k$ are some desities.

In the remainder of this research I assume $M = 1$, so that

$$p(x|\theta; M = 1) = \frac{1}{1 + \lambda\Delta} f_X(x) + \frac{\lambda\Delta}{1 + \lambda\Delta} f_{X+Q}(x), \qquad (5)$$

where $X := \left(\mu - \frac{1}{2}\sigma^2\right)\Delta + \sigma\Delta W_t$, and $Q \sim f_Q$. The first term on the right-hand side of (5) is referred to as the diffusion component, whereas the second one - the jump-diffusion component. The model in which logarithmic rates of return are assumed to follow the distribution given by (5) is further referred to as the DEJD model. In what follows, for simplicity, density (5) is denoted as $p(\cdot|\theta)$ rather than $p(\cdot|\theta; M = 1)$.

Note that for the Kou model ([5]) (Kou model is a special case of the Pareto-Beta Jump-Diffusion specification proposed in ([14])) the density of logarithmic rates of return is given by:

$$p(x) = (1 - \lambda\Delta) f_X(x) + \lambda\Delta f_{X+Q}(x),$$

for $\lambda\Delta < 1$. It is easy to see that

$$p\left(x \middle| \mu, \sigma, \frac{\lambda}{1 + \lambda\Delta}, p_U, \eta_U, \eta_D; \text{Kou}\right) = p(x | \mu, \sigma, \lambda, p_U, \eta_U, \eta_D; \text{DEJD}),$$

and for $\lambda\Delta < 1$

$$p(x | \mu, \sigma, \lambda, p_U, \eta_U, \eta_D; \text{Kou}) = p\left(x \middle| \mu, \sigma, \frac{\lambda}{1 - \lambda\Delta}, p_U, \eta_U, \eta_D; \text{DEJD}\right).$$

A weight ratio of the jump-diffusion and the diffusion weight $\frac{\frac{\lambda\Delta}{1+\lambda\Delta}}{\frac{1}{1+\lambda\Delta}} = \lambda\Delta$ equals the weight ratio $\frac{\exp(-\lambda\Delta)\lambda\Delta}{\exp(-\lambda\Delta)} = \lambda\Delta$ in the original model (2). However, the same is not true for the Kou model. Further considerations are limited only to the DEJD model and the case of daily logarithmic rates of return with $\Delta = \frac{1}{252}$.

## 3  The Bayesian DEJD model

A Bayesian statistical model is defined by the joint density:

$$p(x, \theta) = p(x|\theta) p(\theta),$$

where $x = (x_1, ..., x_n)$ is the observed data, $\theta$ is a vector of unknown parameters, $p(x|\theta)$ is a sampling density and $p(\theta)$ is a prior density. The inference rests upon the posterior density $p(\theta|x)$ of $\theta$ given data $x$ ([27]). If $x_1, ..., x_n$ are mutually independent, then

$$p(\theta|x) = \frac{p(x|\theta)p(\theta)}{p(x)} = \frac{\prod_{i=1}^{n} p(x_i|\theta)p(\theta)}{\int_\Theta \prod_{i=1}^{n} p(x_i|\theta)p(\theta)\,d\theta}.$$

Given $x$, $p(x|\theta)$ — as a function of $\theta$ — is called the likelihood function, whereas

$$p(x) = \int_\Theta \prod_{i=1}^{n} p(x_i|\theta)p(\theta)\,d\theta$$

is the marginal data density, which is invariant with respect to $\theta$, so that

$$p(\theta|x) \propto \prod_{i=1}^{n} p(x_i|\theta)p(\theta).$$

In the present section we set the DEJD model in the Bayesian framework. To facilitate the process, we apply the following reparametrization: $\mu' = \mu - \frac{1}{2}\sigma^2$, $h = \frac{1}{\sigma^2}$, $L = \lambda\Delta$, so that $\theta = \left(\mu', h, L, p_U, \eta_U, \eta_D\right)$. When one analyses a time series which is (or, rather, is believed to be) a trajectory of a jump-diffusion process, then one does not actually know if a given data-point observation has been generated by the pure diffusion or the jump-diffusion component. In other words, one cannot determine which component of the series in (5), i.e. $f_X(x)$ or $f_{X+Q}(x)$ is "responsible for" the observation. To manage the problem let us introduce latent variables $\xi = (\xi_1, ..., \xi_n)$, where $\xi_i \in \{-1, 0, 1\}$ and $P(\xi_i = -1|\theta) = \frac{L}{1+L}p_D$, $P(\xi_i = 0|\theta) = \frac{1}{1+L}$, $P(\xi_i = 1|\theta) = \frac{L}{1+L}p_U$. The value $\xi_i = 0$ means no jump at $t = i\Delta$. The values $\xi_i = -1$ and $\xi_i = 1$ mean that jump occurs and its value is negative or positive, respectively. Moreover, it is convenient to introduce latent variables $J = (J_1, ..., J_n)$ corresponding to the jump value, where

$$p(J_i = j|\theta, \xi_i = -1) = \eta_D \exp(\eta_D j)\,\mathbb{I}_{(-\infty, 0)}(j), \qquad (6)$$
$$p(J_i = j|\theta, \xi_i = 0) = \delta_0,$$
$$p(J_i = j|\theta, \xi_i = 1) = \eta_U \exp(-\eta_U j)\,\mathbb{I}_{(0, \infty)}(j),$$

and $\delta_0$ is the Dirac delta. Then, the $(2n + 6)$-sized vector of all the unknown quantities is denoted by:

$$(\theta, \xi, J) = \left(\mu', h, L, p_u, \eta_u, \eta_d, \xi_1, ..., \xi_n, J_1, ..., J_n\right).$$

Moreover,

$$p(x_i|\theta, \xi_i, J_i) = p(x_i|\theta, J_i) \qquad (7)$$
$$= \frac{1}{\sqrt{2\pi}}\sqrt{\frac{h}{\Delta}}\exp\left(-\frac{1}{2}\frac{h}{\Delta}\left(x_i - \mu'\Delta - J_i\right)^2\right).$$

The Bayesian model is given by:

$$p(x, \theta, \xi, J) = p(x | \theta, \xi, J) p(\theta, \xi, J)$$
$$= p(x | \theta, J) p(\theta, \xi, J).$$

Let the prior structure for $(\theta, \xi, J)$ be defined as:

$$p(\theta, \xi, J) = p(\mu' | h) p(h) p(L) p(p_U) p(\eta_U) p(\eta_D) \cdot$$
$$\cdot \prod_{i=1}^{n} p(J_i | \xi_i, \eta_D, \eta_U, L) \prod_{i=1}^{n} p(\xi_i | p_U, L),$$

where
$p(h) \sim G(v_h, A_h)$ (gamma distribution[1]),
$p(\mu' | h) \sim N(\mu_0, (hA_\mu)^{-1})$ (normal distribution[2]),
$p(L) \sim \chi^2(\gamma_L)$ ($\chi^2$ distribution[3]),
$p(\eta_U) \sim G(\nu_{U,\eta}, A_{U,\eta})$, $p(\eta_D) \sim G(\nu_{D,\eta}, A_{D,\eta})$,
$p(p_U | a_U, b_U) \sim B(a_U, b_U)$ (beta distribution),
$P(\xi = (l_1, ..., l_n) | \theta) = \Pi_{j \in \{-1,0,1\}} w_j^{n_j}$, where $n_j = \#\{i \in \{1, 2, ..., n\} : l_i = j\}$,
$w_{-1} = \frac{L}{1+L} p_D$, $w_0 = \frac{1}{1+L}$, $w_1 = \frac{L}{1+L} p_U$,

$p(J_i = x_i | \theta, \xi_i = -1) \sim -G(1, \eta_d)$,
$p(J_i = x_i | \theta, \xi_i = 1) \sim G(1, \eta_u)$,
$p(J_i = x_i | \theta, \xi_i = 0) = \delta_0$.

Posterior characteristics of the unknown quantities are calculated via the Markov Chain Monte Carlo (MCMC) methods ([28]), combining the Gibbs sampler, the independence and the sequential Metropolis-Hastings algorithms, as well as the acceptance-rejection sampling ([29]). The theorems below make the algorithm ready to use.

**Theorem 1** *Under the above assumptions:*

1. $p\left(\mu', h \middle| x, \theta_{\setminus\{\mu',h\}}, \xi, J\right) \propto$
   $p_G\left(h; n/2 + v_h, \frac{1}{2}\frac{ns}{\Delta} + A_h + \frac{1}{2}\frac{A_\mu n \Delta\left(\mu_0 - \frac{\overline{x} - \overline{J}}{\Delta}\right)^2}{A_\mu + n\Delta}\right) \cdot$
   $\cdot \phi\left(\mu'; \frac{\mu_0 A_\mu + (\overline{x} - \overline{J})n}{A_\mu + n\Delta}, \frac{1}{h(A_\mu + n\Delta)}\right)$

2. $p(L | x, \theta_{\setminus L}, \xi, J) \propto L^{N + \frac{\gamma_L}{2} - 1} \exp\left(-\frac{L}{2}\right) \frac{1}{(1+L)^n}$ where $N = n_{-1} + n_1$, $n_j = \#\{i \in \{1, 2, ..., n\} : l_i = j\}$

3. $p(p_u | x, \theta_{\setminus p_u}, \xi, J) \sim B(n_1 + a_U, n_{-1} + b_U)$

---

[1] $p_G(h; a, b) \propto h^{a-1} \exp(-hb)$ is the density of a gamma distribution.

[2] $N(m, v)$ denotes a normal distribution with mean $m$ and variance $v$.

[3] $p_{\chi^2(\nu)}(x) \propto x^{\frac{\nu}{2} - 1} \exp\left(-\frac{x}{2}\right) \mathbb{I}_{(0,\infty)}(x)$ is the density of a $\chi^2$ distribution with $\nu$ degrees of freedom.

4. $p\left((\eta_D, \eta_U)\,|x, \theta_{\setminus(\eta_D,\eta_U)}, \xi, J\right) \sim$
   $\Gamma\left(\eta_D; (n_{D,\xi} + \nu_{D,\eta}), (A_{D,\eta} - N_{D,J})\right) \cdot$
   $\cdot \Gamma\left(\eta_U; (n_{U,\xi} + \nu_{U,\eta}), (A_{U,\eta} + N_{U,J})\right)$, where $N_{D,J} = \sum_{i=1}^n J_i \mathbb{I}_{(-\infty,0)}(J_i)$,
   $N_{U,J} = \sum_{i=1}^n J_i \mathbb{I}_{(0,\infty)}(J_i)$,

5. $p(\xi, J\,|x, \theta) = \prod_{i=1}^n p(J_i\,|x_i, \theta, \xi_i) p(\xi_i\,|x_i, \theta)$, where

   (a) $P(\xi_i = 0|x_i, \theta) = \frac{1}{G} \frac{1}{\sigma\sqrt{\Delta}} \phi\left(\frac{x_i - \mu'\Delta}{\sigma\sqrt{\Delta}}; 0, 1\right)$

   (b) $P(\xi_i = -1|x_i, \theta) = \frac{1}{G} \eta_D \exp\left(\eta_D x_i - \mu'\Delta\eta_D + \frac{1}{2}\sigma^2\Delta\eta_D^2\right) \cdot$
   $\cdot \Phi\left(-\frac{x_i - (\mu'\Delta - \sigma^2\Delta\eta_D)}{\sigma\sqrt{\Delta}}; 0, 1\right) Lp_D$

   (c) $P(\xi_i = 1|x_i, \theta) = \frac{1}{G} \eta_U \exp\left(-\eta_U x_i + \mu'\Delta\eta_U + \frac{1}{2}\sigma^2\Delta\eta_U^2\right) \cdot$
   $\cdot \Phi\left(\frac{x_i - (\mu'\Delta + \sigma^2\Delta\eta_U)}{\sigma\sqrt{\Delta}}; 0, 1\right) Lp_U$

   (d) $p(J_i = j\,|x_i, \theta, \xi_i = 0) = \delta_0(j)$,

   (e) $p(J_i = j\,|x_i, \theta, \xi_i = -1) \propto \phi\left(j; x_i - \mu'\Delta + \frac{\Delta}{h}\eta_D, \frac{\Delta}{h}\right) \mathbb{I}_{(-\infty,0)}(j)$

   (f) $p(J_i = j\,|x_i, \theta, \xi_i = 1) \propto \phi\left(j; x_i - \mu'\Delta - \frac{\Delta}{h}\eta_U, \frac{\Delta}{h}\right) \mathbb{I}_{(0,\infty)}(j)$

   and
   $G := \frac{1}{\sigma\sqrt{\Delta}} \phi\left(\frac{x_i - \mu'\Delta}{\sigma\sqrt{\Delta}}; 0, 1\right) + \eta_D \exp\left(\eta_D x_i - \mu'\Delta\eta_D + \frac{1}{2}\sigma^2\Delta\eta_D^2\right) \cdot$
   $\cdot \Phi\left(-\frac{x_i - (\mu'\Delta - \sigma^2\Delta\eta_D)}{\sigma\sqrt{\Delta}}; 0, 1\right) \cdot Lp_D +$
   $+ \eta_U \exp\left(-\eta_U x_i + \mu'\Delta\eta_U + \frac{1}{2}\sigma^2\Delta\eta_U^2\right) \Phi\left(\frac{x_i - (\mu'\Delta + \sigma^2\Delta\eta_U)}{\sigma\sqrt{\Delta}}; 0, 1\right) \cdot Lp_U$.

The Gibbs algorithm rests upon sampling from the full conditional distributions. Since $p\left(\mu', h\,|x, \theta_{\setminus\{\mu',h\}}, \xi, J\right)$, $p\left(\eta_D, \eta_U\,|x, \theta_{\setminus(\eta_D,\eta_U)}, \xi, J\right)$ and $p\left(p_u\,|x, \theta_{\setminus p_u}, \xi, J\right)$ are densities of the gamma-normal, gamma and beta distributions, sampling $\mu'$, $h$, $p_u$, $\eta_D$ and $\eta_U$ is straightforward. Generating latent variabes $\xi_i$ for $i = 1, ..., n$ does not pose a challenge either, as for each $i$ variable $\xi_i$ (given $x_i$ and $\theta$) has a discrete distribution with probabilities given in Theorem 1. Also generating $J_i$ under given $x_i$, $\theta$ and $\xi_i = -1$ or $\xi_i = 1$, is easy because the distributions are truncated normal distributions. Note that if $\xi_i = 0$, then $J_i \equiv 0$. Sampling from $p(L\,|x, \theta_{\setminus L}, \xi, J)$ is managed according to the following alternative propositions.

**Proposition 2** 1. The independent Metropolis-Hastings algorithm with the candidate-generating density $(2n+1)L \sim \chi^2_{2N+\nu_L}$ and the transition probability:

$$\min\left\{\exp\left(n\left(L^{(m+1)} - L^{(m)}\right)\right)\left[1 + L^{(m+1)}\right]^{-n}\left[1 + L^{(m)}\right]^n, 1\right\},$$

from a state $L^{(m)}$ to $L^{(m+1)}$ can be used to sample from $p(L\,|x, \theta_{\setminus L}, \xi, J)$.

2. If $n - N - \frac{\nu_L}{2} > 0$, then the acceptance-rejection sampling with a proposition density of the gamma-gamma distribution [4][5]:

$$L \sim Gg\left(L; n - N - \frac{\nu_L}{2}, 1, N + \frac{\nu_L}{2}\right),$$

and the acceptance probability $e^{-L/2}$ can be used to sample from $p\left(L \mid x, \theta_{\setminus L}, \xi, J\right)$.

# 4 Examples

In this section, we illustrate the methodology developed in the previous section. First, the estimation results of the DEJD model parameters for two simulated time series are presented. Subsequently, the real-world dataset of logarithmic rates of return on KGHM is fit with the DEJD structure.

All the calculations are performed in R. Numerical algorithms applied in the research require monitoring convergence of the generated chain to its limiting stationary distribution. Convergence of all the MCMC samplers exploited in our research is confirmed by visual inspection of the ergodic means, standard deviations and CUMSUM statistics plots ([30]). The results seem to be robust to the choice of the starting point for the MCMC procedure.

In what follows, four different prior structures are considered, with the hyperparameters of each being displayed in Table 1. Formally, each prior specification defines a different Bayesian model, what yields the four $\text{DEJD}_I$, $\text{DEJD}_{II}$, $\text{DEJD}_{III}$, $\text{DEJD}_{IV}$.

## 4.1 Simulation case studies

A series of $n = 10,000$ data points generated from the DEJD process is under consideration. Table 2 presents posterior means and standard deviations along with the true values of the parameters. The presented results are based on $100,000$ MCMC draws, preceeded by $100,000$ burn-in cycles.

The posterior means of $\text{DEJD}_I$ parameters are close to the true values. The posterior expectations of $\mu$, $\eta_U$ and $\lambda$, calculated under $\text{DEJD}_{II}$, differ substantially from the prespecified values. Values $E\left(\frac{1}{\eta_U} \mid x\right)$ and $E\left(\frac{1}{\eta_D} \mid x\right)$ are the posterior means of negative and positive jumps values, respectively. The value of $E\left(\frac{1}{\eta_U} \mid x\right)$ calculated under prior $II$ is greater than the one obtained under prior $I$, so the positive jumps are (on average) greater under prior $II$. Note that the probability of positive jumps $p_U$ and the jump intensity $\lambda$ are greater for $\text{DEJD}_I$. Hence, one can expect that the number of detected positive jumps is lower for prior $II$, so the role of the jump component, under the $\text{DEJD}_{II}$ framework, is smaller than in the case of prior $I$. It seems to be supported by a greater value of the trend parameter, $\mu$, in $\text{DEJD}_{II}$. Figure 1 displays the marginal posteriors of parameters in the $\text{DEJD}_I$ model, along with the prior densities. The prior distributions of $\eta_D$ and $\eta_U$ allow for large values of parameters. The data move the posterior probability to the left of the prior mode and the posterior means of the parameters stay close to the true values specified for $\eta_D$ and $\eta_U$.

---

[4] In practice, the condition $n - N - \frac{\nu_L}{2} > 0$ is often satisfied.

[5] $x \to \frac{\beta^\alpha}{\Gamma(\alpha)} \frac{\Gamma(\alpha+n)}{\Gamma(n)} \frac{x^{n-1}}{(\beta+x)^{\alpha+n}} \mathbb{I}_{(0,\infty)}(x)$, where $\alpha > 0, \beta > 0, n = 1, 2, ...$ is the density of gamma-gamma distribution $Gg(x; \alpha, \beta, n)$ ([27]).

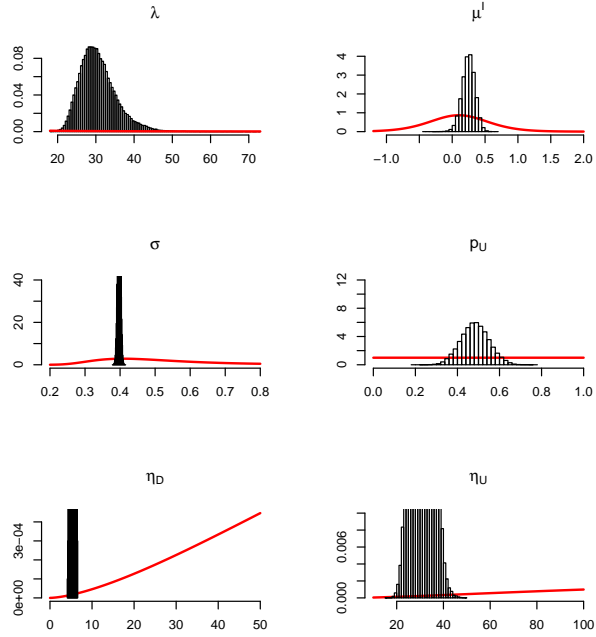

Figure 1: Marginal posterior (bars) and prior densities (solid line) of parameters in the DEJD$_I$ model.

Figure 2 displays the marginal posteriors of parameters in the DEJD$_{II}$ model, along with the prior densities. The plots reveal a considerable contribution of the data to the shape of marginal posteriors. The prior specifications for $\eta_D$ and $\eta_U$ support lower values of the parameters. However, the data move the posterior distributions to the right (into the right tails) towards the true values.

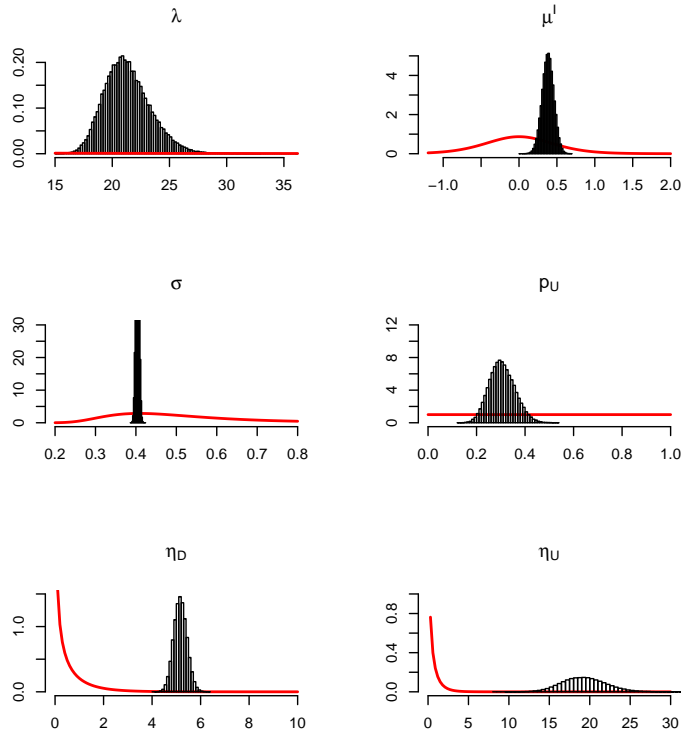

Figure 2: Marginal posterior (bars) and prior densities (solid line) of parameters in the DEJD$_{II}$ model.

Only in the case of prior $I$, the data were strong enough to move the posteriors of $\eta_D$ and $\eta_U$ close to the true values. We observe the impact of the inverse gamma prior distribution parameters upon the posterior distribution. The parameter prior impact on the posterior results was also observed in the case of a normal jump distribution (the JD(M)J model, [23]).

## 4.2 Analysis for the KGHM

KGHM is a copper producer and one of the largest Polish exporters. The company's share prices contribute to the WIG20 Index. Let us consider a time series of the daily values of logarithmic growth rates of KGHM quotations on the Warsaw Stock Exchange from January 23, 2006 to February 22, 2010.

I analysed many prior specifications and rejected models with too high intensity of jumps (e.g. jump at every day), according to the idea of a jump as a sporadic event. Moreover, I focused on priors which correspond to data, i.e. I rejected the ones under which the histograms of posterior marginals were situated in very low prior probability regions (far in the priors' tails). Finally, I chose prior $III$. Table 3 contains basic posterior characteristics of the DEJD$_{III}$ model's parameters. The presented results are based on $70,000$ MCMC draws, preceeded by $30,000$ burn-in cycles.

The value $E(\lambda|x) = 26.5479$ indicates that jumps arrive every $9-10$ days. Let us remind that the probability of the pure diffusion component equals $\frac{1}{1+\lambda\Delta}$ at each period of time $\Delta$ (cf. (5)). The posterior mean $E\left(\frac{1}{1+\lambda\Delta}|x\right) = 0.9062$ implies that this probability is high and the frequency of jumps is low (which coincides with our expectations). Figure 3 displays the marginal posterior and prior densities of parameters under $\text{DEJD}_{III}$. Note that $E(p_U|x) = 0.2863$ is less than 0.5 and the marginal posterior distribution of $p_U$ has its probability mass shifted to the left of 0.5, so negative jumps are more probable a posteriori than the positive ones.

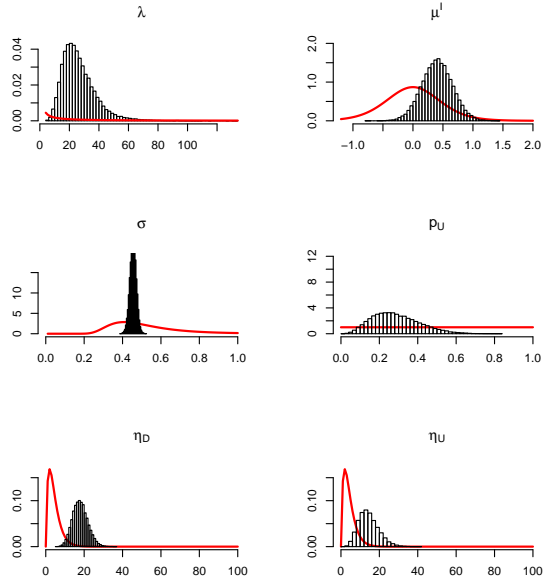

Figure 3: Marginal posterior (bars) and prior densities (solid line) of parameters in the $\text{DEJD}_{III}$ model estimated for the log-returns on the KGHM share prices.

Analysing the posterior means and the marginal posterior distributions of $1/\eta_D$ and $1/\eta_U$ it might be concluded that the absolute values of the negative and the positive jumps are similar. However, note that if we assume that parameters of the model equal to the posterior means then the density $Q_j$ (cf. (1)), depicted in Figure 4, is asymmetrical. This is due to the inequality between $p_D$ and $p_U$. This observation supports the application of the asymmetrical jump distribution and the DEJD specification to the data at hand.

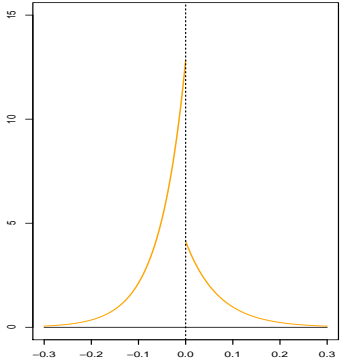

Figure 4: The density of $Q_j$ with parameters equal to the posterior means under the DEJD$_{III}$ model.

**Jump detection**

Jumps in prices or returns are often defined as the values exceeding some arbitrarily choosen thresholds. Different thresholds lead to a different number of jumps ([16]). Thresholds are commonly set symmetrically either around zero or the sample mean and are defined as a multiply of the sample standard deviation. In many cases, the empirical distribution of logarithmic rates of return features negative skewness. Hence, symmetric thresholds do not seem valid then. In what follows, the latent variables $\xi_i$ are used to identify data points with a jump.

As mentioned before, when one analyses a time series which is believed to be a trajectory of a DEJD process, then one does not know if a given datapoint observation has been generated by the pure diffusion or the jump-diffusion component. Formally, an event of jump appearance is equivalent to $\xi_i = -1$ or $\xi_i = 1$. Unfortunately, we do not observe $\xi_i$, but we can assess the posterior probability of a jump: $P(\xi_i \neq 0 | x)$ for each day $i = 1, ..., n$. Let us assume that a jump occurs at the $i$-th period if probability $P(\xi_i \neq 0 | x)$ exceeds an arbitrarily choosen value of 0.5, which corresponds to the aforementioned thresholds. However, the problem of asymmetry or symmetry is not a matter here.

Let $J_D$ and $J_U$ denote the smallest (in terms of absolute value) logarithmic rates of return for which the posterior probability of a jump exceeds the prespecified value of 0.5. Define $k_D = \frac{\overline{x}_n - J_D}{\sigma_n}$ and $k_U = \frac{J_U - \overline{x}_n}{\sigma_n}$, where $\overline{x}_n$ and $\sigma_n$ equal the sample mean and the standard deviation, respectively. Then $k_D = 2.22$ and $k_U = 2.66$. Figure 5 depicts the modeled time series (with a band of $\overline{x}_n - k_D \cdot \sigma_n$ and $\overline{x}_n + k_U \cdot \sigma_n$), and the values of probabilities $P(\xi_i = -1 | x)$ and $P(\xi_i = 1 | x)$ against the number of successive days, provided that $P(\xi_i = -1 | x) > 0.5$ and $P(\xi_i = 1 | x) > 0.5$. Note that higher posterior probabilities of a jump go along with higher volatility of the time series. Moreover, there are more negative jumps than the positive ones and the interval $[\overline{x}_n - k_D \cdot \sigma_n, \overline{x}_n + k_U \cdot \sigma_n]$ is not symmetric around the sample mean.

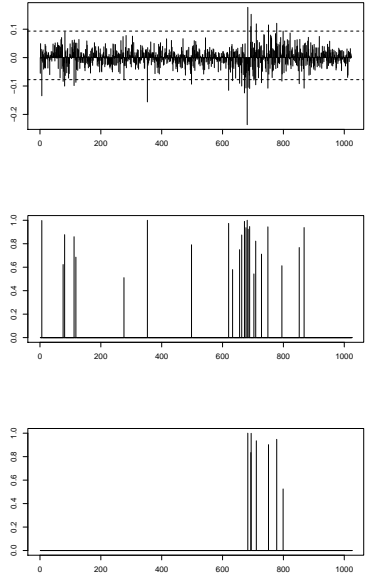

Figure 5: The modeled time series of daily log-returns (top),
$i \to P(\xi_i = -1 \,|x) \mathbb{I}_{\{i:P(\xi_i=-1|x)>0.5\}}(i)$ (middle), and
$i \to P(\xi_i = 1 \,|x) \mathbb{I}_{\{i:P(\xi_i=1|x)>0.5\}}(i)$ (bottom).

Noteworthy, periods of no jumps alternate with the ones of frequent jumps, which hints at the existence of jump clustering (the same conclusion was drawn under the JD($M$)J specifications in [22]). Although for many days the posterior probability of a jump is fairly high, most of periods feature low probabilities. This observation justifies importance of the pure-diffusion component.

Frame and Ramezani analysed the Bayesian Bernoulli jump-diffusion model with a double exponential jump distribution. Their prior specification was based on non-informative priors (with an exception of $\lambda$). They modeled logarithmic rates of return of the S&P500 index. Their results[6] indicate "high" intensity ($E(\lambda|x) \approx 722$) of "small" jumps ($E(\eta_D|x) \approx 60.5$, $E(\eta_U|x) \approx 62$), which does not correspond to the idea of seldom jumps and extreme values. The DEJD$_{IV}$ model provides similar assessments for the S&P500 index. Different priors lead to various Bayesian models and they may yield different posteriors of the parameters. Under such circumstances the interpretation of the mathematical model parameters' estimates is not straightforward, though the Bayesian pooling approach may prove of merit in the context ([31]).

From the definitions of the aforementioned mathematical jump-diffusion models and their discretizations it follows that the probability of small absolute values of jumps is not equal zero (cf. Figure 4). This may be attributed to modeling jumps by mean of the normal and the double exponential distributions. Hence, the absolute values of jumps might be close to zero. This fact is in contradiction to the notation of jumps as extreme values. The role of the jump component should be different from the pure diffusion one. As was mentioned above, the results of the Bayesian analyses depend on the priors. Otherwise, the priors' hyperparameters yield, in an elegant and formal way, a tool to separate jumps from the pure diffusion. This ability suppresses the drawback of the mathematical model and it shows the power of the Bayesian inference. It appears that it is the mathematical (rather than the statistical) model that is the reason behind the ambiguities in the results of estimation.

---

[6]The results from p. 25 Table 5, S&P500 05/2007-03/2009 (Bear Market) from [21].

## 5 Conclusions

In the paper, the Bayesian DEJD model is developed. To employ the model in practice numerical techniques based on the MCMC methods are proposed. The Bayesian statistics equipped with the MCMC mehods gives us an easy way of estimating parameters of the DEJD model. The methodology is illustrated with a simulation experiment and an empirical study, in which logarithmic rates of return on the KGHM share are analysed. Latent variables enable detection of negative and positive jumps and analysis of their frequency and distributions. The empirical results support applications of the jump-diffusion models with an asymmetric distribution of jumps. Periods of no jumps and the ones of frequent jumps suggest the existence of jump clustering. Unfortunately, the results might hinge on the prior assumptions. This feature is commonly observed in models based on mixture distributions ([32], [33]).

In the Merton model, the JD($M$)J model and the Bernoulli diffusion model the jump value distributions are normal. If a mean of normal distribution is not equal zero, than the distribution is asymmetric with respect to zero, so the jump-diffusion model with a double exponential distribution and its discrete approximations, such as the DEJD model, constitutes only an alternative tool of modeling asymmetric jumps and not their generalization.

The proposed Bayesian DEJD model might be employed in VaR analyses and in pricing derivative securities.

*Table 1: Priors structures*

| Priors | $\mu^{'}$ | $A_\mu$ | $\nu_h$ | $A_h$ | $A_U$ | $B_U$ | $\upsilon_{\eta_U}$ | $A_{\eta_U}$ | $\upsilon_{\eta_D}$ | $A_{\eta_D}$ | $\nu_L$ |
|---|---|---|---|---|---|---|---|---|---|---|---|
| I | 0.1 | 1 | 5 | 1 | 1 | 1 | 2.56 | 0.00576 | 2.56 | 0.00576 | $10\Delta$ |
| II | 0 | 1 | 5 | 1 | 1 | 1 | 0.5 | 1 | 0.5 | 1 | $10\Delta$ |
| III | 0 | 1 | 5 | 1 | 1 | 1 | 1.86 | 0.43 | 1.86 | 0.43 | $10\Delta$ |
| IV | 0 | 1 | 2.56 | 0.00576 | 1 | 1 | 2.56 | 0.00576 | 2.56 | 0.00576 | $10\Delta$ |

*Table 2: Posterior means and standard deviations for simulation data and the true parameters of the model.*

| Model | $\text{DEJD}_I$ | | $\text{DEJD}_{II}$ | | True |
|---|---|---|---|---|---|
| $\theta$ | $E(\cdot|x)$ | $D(\cdot|x)$ | $E(\cdot|x)$ | $D(\cdot|x)$ | $\theta$ |
| $\mu$ | 0.3262 | 0.0973 | 0.4632 | 0.0781 | 0.25 |
| $\sigma$ | 0.3972 | 0.0043 | 0.4039 | 0.0039 | 0.4 |
| $p_U$ | 0.4835 | 0.0666 | 0.3055 | 0.0526 | 0.5 |
| $\eta_D$ | 5.3202 | 0.2778 | 5.1647 | 0.2721 | 5 |
| $\eta_U$ | 30.6779 | 3.7369 | 19.2997 | 2.6807 | 30 |
| $\lambda$ | 30.6419 | 4.8556 | 21.3400 | 2.0318 | 30 |

*Table 3: Posterior means and standard deviations for KGHM.*

| | $\text{DEJD}_{III}$ | |
|---|---|---|
| $\theta$ | $E(\cdot|x)$ | $D(\cdot|x)$ |
| $\mu^{'}$ | 0.4038 | 0.2559 |
| $\mu$ | 0.5074 | 0.2553 |
| $\sigma$ | 0.4548 | 0.0168 |
| $\lambda$ | 26.5479 | 11.5128 |
| $\frac{1}{1+\lambda\Delta}$ | 0.9062 | 0.0356 |
| $p_U$ | 0.2863 | 0.1215 |
| $\eta_D$ | 17.9707 | 3.9736 |
| $\eta_U$ | 14.4137 | 5.1729 |
| $1/\eta_D$ | 0.0586 | 0.0142 |
| $1/\eta_U$ | 0.0801 | 0.0356 |

# 6  Appendix

**Lemma 1** *Under the conditions stated in the paper, the likelihood function is given by*

$$p(x \mid \theta, \xi, J) = h^{n/2} \exp\left(-\frac{1}{2}h\left\{\frac{ns}{\Delta} + n\Delta\left(\frac{\overline{x}-\overline{J}}{\Delta} - \mu'\right)^2\right\}\right), \quad (8)$$

*where*

$$s = \frac{1}{n}\sum_{i=1}^{n}\left(x_i - J_i - (\overline{x}-\overline{J})\right)^2,$$

$$\overline{x} = \frac{1}{n}\sum_{i=1}^{n} x_i, \quad \overline{J} = \frac{1}{n}\sum_{i=1}^{n} J_i.$$

**Lemma 2** *Under the conditions stated in the paper,*

1. 
$$A_\mu\left(\mu_0 - \mu'\right)^2 + n\Delta\left(\frac{\overline{x}-\overline{J}}{\Delta} - \mu'\right)^2 = \frac{A_\mu n\Delta\left(\mu_0 - \frac{\overline{x}-\overline{J}}{\Delta}\right)^2}{A_\mu + n\Delta}. \quad (9)$$

2. 
$$\int_{-\infty}^{\infty} \frac{1}{\sigma\sqrt{\Delta}} \phi\left(\frac{z - \mu'\Delta}{\sigma\sqrt{\Delta}}; 0, 1\right) \eta_D \exp(\eta_D(x-z)) \mathbb{I}_{(-\infty, 0)}(x-z)\, dz$$
   (10)
$$= \eta_D \exp\left(\eta_D x - \mu'\Delta\eta_D + \frac{1}{2}\sigma^2\Delta\eta_D^2\right) \Phi\left(-\frac{x - \left(\mu'\Delta - \sigma^2\Delta\eta_D\right)}{\sigma\sqrt{\Delta}}; 0, 1\right),$$

   *where $\phi(\cdot; m, v)$ and $\Phi(\cdot; m, v)$ are the density and the cumulative distribution of the normal distribution $N(m, v)$, respectively.*

3. 
$$\int_{-\infty}^{\infty} \frac{1}{\sigma\sqrt{\Delta}} \phi\left(\frac{z - \mu'\Delta}{\sigma\sqrt{\Delta}}; 0, 1\right) \eta_U \exp(-\eta_U(x-z)) \mathbb{I}_{[0,\infty)}(x-z)\, dz \quad (11)$$
$$= \eta_U \exp(-\eta_U x) \exp\left(\mu'\Delta\eta_U + \frac{1}{2}\sigma^2\Delta\eta_U^2\right) \Phi\left(\frac{x - \left(\mu'\Delta + \sigma^2\Delta\eta_U\right)}{\sigma\sqrt{\Delta}}; 0, 1\right).$$

4. 
$$\sqrt{\frac{h}{\Delta}} \phi\left(\sqrt{\frac{h}{\Delta}}\left(x_i - \mu'\Delta - j\right); 0, 1\right) \eta_D \exp(\eta_D j) \mathbb{I}_{(-\infty, 0)}(j) \quad (12)$$
$$= C \exp\left(-\frac{1}{2}\frac{h}{\Delta}\left(j - \left[(x_i - \mu'\Delta) + \frac{\Delta}{h}\eta_D\right]\right)^2\right) \mathbb{I}_{(-\infty, 0)}(j),$$

   *where $C$ does not depend on $j$.*

5.

$$\sqrt{\frac{h}{\Delta}}\phi\left(\sqrt{\frac{h}{\Delta}}\left(x_i - \mu'\Delta - j\right);0,1\right)\eta_U \exp(-\eta_U j)\,\mathbb{I}_{(0,\infty)}(j) \quad (13)$$

$$= C\exp\left(-\frac{1}{2}\frac{h}{\Delta}\left[\left(j - \left[\left(x_i - \mu'\Delta\right) - \frac{\Delta}{h}\eta_U\right]\right)^2\right]\right)\mathbb{I}_{(0,\infty)}(j),$$

where $C$ does not depend on $j$.

**Proof.** Tedious, but simple calculations lead to the claims. ∎

**Theorem 1** *Under the conditions stated in the paper,*

1. $p\left(\mu', h \mid x, \theta_{\setminus\{\mu',h\}}, \xi, J\right) \propto$
   $p_G\left(h; n/2 + \upsilon_h, \frac{1}{2}\frac{ns}{\Delta} + A_h + \frac{1}{2}\frac{A_\mu n\Delta\left(\mu_0 - \frac{\bar{x}-\bar{J}}{\Delta}\right)^2}{A_\mu + n\Delta}\right) \cdot$
   $\cdot \phi\left(\mu'; \frac{\mu_0 A_\mu + (\bar{x}-\bar{J})n}{A_\mu + n\Delta}, \frac{1}{h(A_\mu + n\Delta)}\right)$

2. $p\left(L \mid x, \theta_{\setminus L}, \xi, J\right) \propto L^{N+\frac{\gamma_L}{2}-1}\exp\left(-\frac{L}{2}\right)\frac{1}{(1+L)^n}$,
   where $N = n_{-1} + n_1$, $n_j = \#\{i \in \{1,2,...,n\}: l_i = j\}$

3. $p\left(p_u \mid x, \theta_{\setminus p_u}, \xi, J\right) \sim B(n_1 + a_U, n_{-1} + b_U)$

4. $p\left((\eta_D, \eta_U) \mid x, \theta_{\setminus(\eta_D,\eta_U)}, \xi, J\right) \sim$
   $\Gamma\left(\eta_D; (n_{D,\xi} + \nu_{D,\eta}), (A_{D,\eta} - N_{D,J})\right) \cdot$
   $\cdot \Gamma\left(\eta_U; (n_{U,\xi} + \nu_{U,\eta}), (A_{U,\eta} + N_{U,J})\right)$, where $N_{D,J} = \sum_{i=1}^n J_i \mathbb{I}_{(-\infty,0)}(J_i)$,
   $N_{U,J} = \sum_{i=1}^n J_i \mathbb{I}_{(0,\infty)}(J_i)$,

5. $p(\xi, J \mid x, \theta) = \prod_{i=1}^n p(J_i \mid x_i, \theta, \xi_i) p(\xi_i \mid x_i, \theta)$, where

    (a) $P(\xi_i = 0 \mid x_i, \theta) = \frac{1}{G}\frac{1}{\sigma\sqrt{\Delta}}\phi\left(\frac{x_i - \mu'\Delta}{\sigma\sqrt{\Delta}};0,1\right)$

    (b) $P(\xi_i = -1 \mid x_i, \theta) = \frac{1}{G}\eta_D \exp\left(\eta_D x_i - \mu'\Delta\eta_D + \frac{1}{2}\sigma^2\Delta\eta_D^2\right)\cdot$
    $\cdot \Phi\left(-\frac{x_i - (\mu'\Delta - \sigma^2\Delta\eta_D)}{\sigma\sqrt{\Delta}};0,1\right) L p_D$

    (c) $P(\xi_i = 1 \mid x_i, \theta) = \frac{1}{G}\eta_U \exp\left(-\eta_U x_i + \mu'\Delta\eta_U + \frac{1}{2}\sigma^2\Delta\eta_U^2\right)\cdot$
    $\cdot \Phi\left(\frac{x_i - (\mu'\Delta + \sigma^2\Delta\eta_U)}{\sigma\sqrt{\Delta}};0,1\right) L p_U$

    (d) $p(J_i = j \mid x_i, \theta, \xi_i = 0) = \delta_0(j)$,

    (e) $p(J_i = j \mid x_i, \theta, \xi_i = -1) \propto \phi\left(j; x_i - \mu'\Delta + \frac{\Delta}{h}\eta_D, \frac{\Delta}{h}\right)\mathbb{I}_{(-\infty,0)}(j)$

    (f) $p(J_i = j \mid x_i, \theta, \xi_i = 1) \propto \phi\left(j; x_i - \mu'\Delta - \frac{\Delta}{h}\eta_U, \frac{\Delta}{h}\right)\mathbb{I}_{(0,\infty)}(j)$

    and
    $G := \frac{1}{\sigma\sqrt{\Delta}}\phi\left(\frac{x_i - \mu'\Delta}{\sigma\sqrt{\Delta}};0,1\right) + \eta_D \exp\left(\eta_D x_i - \mu'\Delta\eta_D + \frac{1}{2}\sigma^2\Delta\eta_D^2\right)\cdot$

$$\cdot \Phi\left(-\frac{x_i - \left(\mu' \Delta - \sigma^2 \Delta \eta_D\right)}{\sigma\sqrt{\Delta}}; 0, 1\right) Lp_D$$

$$+\eta_U \exp\left(-\eta_U x_i + \mu' \Delta \eta_U + \tfrac{1}{2}\sigma^2 \Delta \eta_U^2\right) \Phi\left(\frac{x_i - \left(\mu' \Delta + \sigma^2 \Delta \eta_U\right)}{\sigma\sqrt{\Delta}}; 0, 1\right) \cdot Lp_U.$$

**Proof.**
■

**Proposition 1**  1. The independent Metropolis-Hastings algorithm with the candidate-generating density $(2n+1)L \sim \chi^2_{2N+\nu_L}$ and the transition probability:

$$\min\left\{\exp\left(n\left(L^{(m+1)} - L^{(m)}\right)\right)\left[1 + L^{(m+1)}\right]^{-n}\left[1 + L^{(m)}\right]^n, 1\right\},$$

from a state $L^{(m)}$ to $L^{(m+1)}$ can be used to sample from $p\left(L \,\middle|\, x, \theta_{\setminus L}, \xi, J\right)$.

2. If $n - N - \frac{\nu_L}{2} > 0$, then the acceptance-rejection sampling with a proposition density of the gamma-gamma distribution [7]:

$$L \sim Gg\left(L; n - N - \frac{\nu_L}{2}, 1, N + \frac{\nu_L}{2}\right),$$

and the acceptance probability $e^{-L/2}$ can be used to sample from $p\left(L \,\middle|\, x, \theta_{\setminus L}, \xi, J\right)$.

**Proof.**
■

---

[7] $x \to \frac{\beta^\alpha}{\Gamma(\alpha)} \frac{\Gamma(\alpha+n)}{\Gamma(n)} \frac{x^{n-1}}{(\beta+x)^{\alpha+n}} \mathbb{I}_{(0,\infty)}(x)$, where $\alpha > 0, \beta > 0, n = 1, 2, ...$ is the density of gamma-gamma distribution $Gg(x; \alpha, \beta, n)$ ([27]).